\documentclass[runningheads]{llncs}
\usepackage{graphicx}
\usepackage{hyperref} 
\usepackage{float} 
\usepackage[firstpage=true, placement=bottom, color=black]{background} 
\usepackage{tikz}
\newcommand\copyrighttext{%
  \footnotesize Accepted at International Conference on Database Systems for Advanced Applications (DASFAA 2019).
  The final authenticated version is available online at: \url{https://doi.org/10.1007/978-3-030-18590-9\_80}.
  }
\newcommand\copyrightnotice{%
\begin{tikzpicture}[remember picture, overlay]
\node[anchor=south, yshift=10pt] at (current page.south) {\fbox{\parbox{\dimexpr\textwidth-\fboxsep-\fboxrule\relax}{\copyrighttext}}};
\end{tikzpicture}%
}
\backgroundsetup{opacity=1, scale=1, angle=0, contents={%
\copyrightnotice
}}

\usepackage{enumitem}
\begin{document}
\title{Adding Value by Combining Business and Sensor Data: An Industry 4.0 Use Case}
\titlerunning{Adding Value by Combining Business and Sensor Data}
%
\author{Guenter Hesse\orcidID{0000-0002-7634-3021} \and
Christoph Matthies\orcidID{0000-0002-6612-5055} \and 
Werner Sinzig \and
Matthias Uflacker}

\authorrunning{G. Hesse et al.}
%

\institute{Hasso Plattner Institute \\ 
August-Bebel-Str. 88, 
14482 Potsdam, Germany \\
\email{\{first name.name\}@hpi.de}}
\maketitle              
\begin{abstract}
Industry 4.0 and the Internet of Things are recent developments that have lead to the creation of new kinds of manufacturing data.
Linking this new kind of sensor data to traditional business information is crucial for enterprises to take advantage of the data's full potential.
In this paper, we present a demo which allows experiencing this data integration, both vertically between technical and business contexts and horizontally along the value chain.
The tool simulates a manufacturing company, continuously producing both business and sensor data, and supports issuing ad-hoc queries that answer specific questions related to the business.
In order to adapt to different environments, users can configure sensor characteristics to their needs.

\end{abstract}
\section{Introduction}
\label{sec:introduction}
The developments in the areas of Internet of Things (IoT) and sensor technologies drive advances in modern manufacturing settings.
Industrial manufacturing enterprises recognize this technological progress and are using the new Industry 4.0 capabilities to generate added value.
For example, on a daily basis, a single sensor located on a General Electric gas turbine engine can produce 500GB of data~\cite{davenport}.
Injection molding machines, as an example of a common manufacturing device, can even generate multiple terabytes of sensor data per day~\cite{DBLP:conf/gi/HuberVN16}.

However, these new possibilities also pose unique challenges, e.g., regarding data integration, as the characteristics of IoT and business data differ~\cite{DBLP:conf/tpctc/HesseRMLKU17}.
Linking these two kinds of data holds the key for unlocking the full potential that lies within the collected data treasure.
Contrary to horizontal integration, which describes the integration of business data along the value chain, vertical integration refers to the connection between business and sensor data.
Whereas in horizontal integrations only homogenous business data needs to be merged, vertical integration requires integration of a variety of data characteristics.

In the presented demo that is available online~\footnote{https://github.com/Gnni/DemoDataIntegration}, we tackle the challenges of understanding the complex data relations in an Industry 4.0 setting.
We present an approach for simulating different types and amounts of sensors in the context of an industrial manufacturing company, which also produces business data as part of its regular activities.
Furthermore, we enable issuing ad-hoc queries on the collected data in an easy-to-use fashion by employing SQL.
This flexibility allows analyzing and combining all kinds of available data.
This allows for horizontal as well as vertical integration in this synthetic and configurable scenario.

\section{Developed Demo System}
\label{sec:developedsystem}

The system is developed using the Play framework~\cite{playframework} and Scala as the programming language.
The demo is realized as a single page application (SPA) which allows controlling data generation for both, business data and sensor data~\cite{Mikowski:2013:SPW:2663433}.
As the application is preconfigured with the default settings of a fictional engine producing factory employing IoT sensors, it can immediately be run and explored.
However, the number and characteristics of sensors that are sending data can be adapted using on-screen controls.
Two live-updating line charts visualize the data ingestion rate for both kinds of data.
This data is inserted into a columnar in-memory database to enable real-time query execution.

\begin{figure}[H]
\center
	\includegraphics[width=0.881\textwidth]{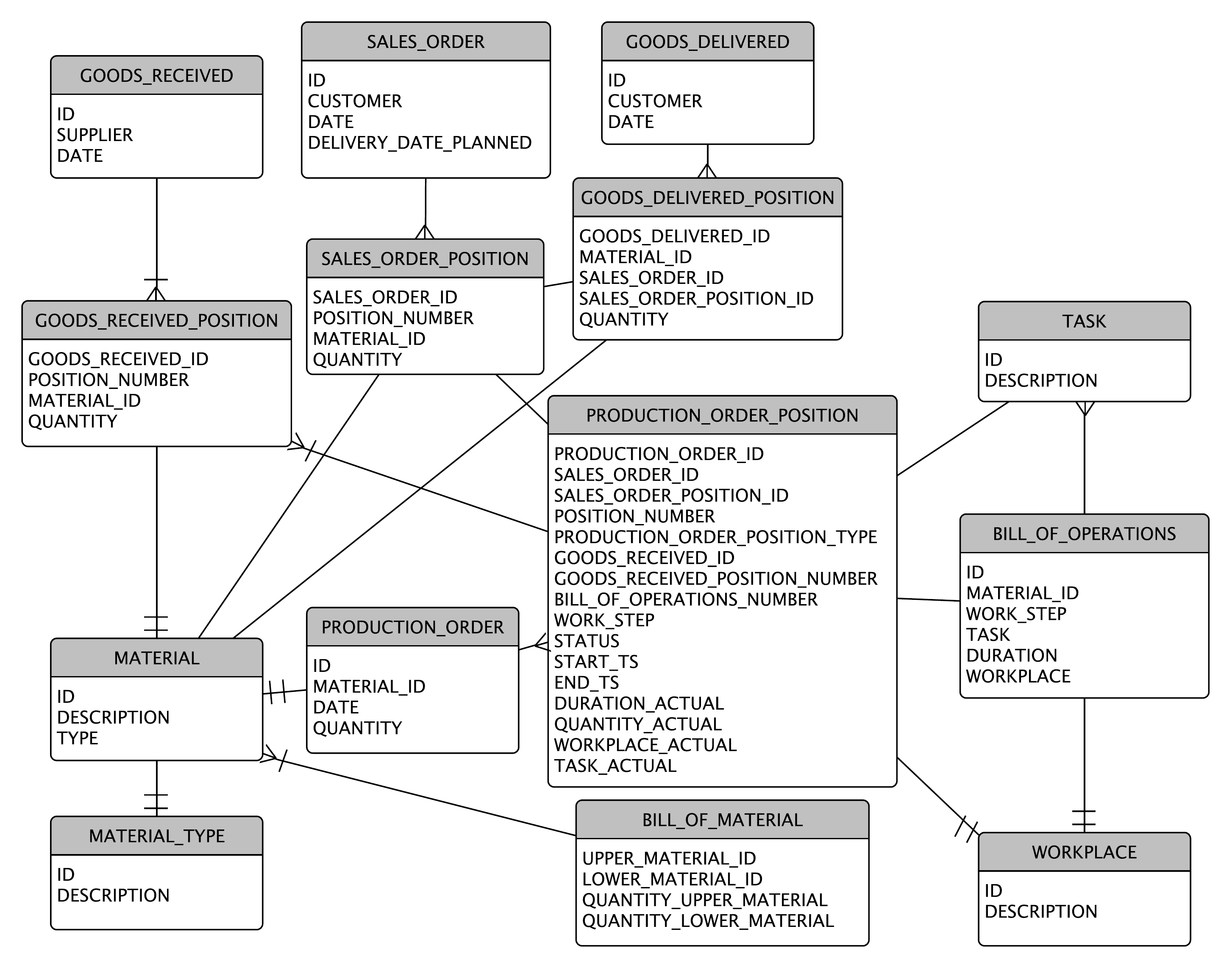}
	\caption{Entity Relationship Diagram in Crow's Foot Notation for the Business Data}
	\label{fig:erd}
\end{figure}

The used data model visualized in Figure~\ref{fig:erd} is inspired by the schemas of real Enterprise Resource Planning (ERP) systems~\cite{DBLP:conf/sigmod/Plattner09}.
Particularly, the idea of having a head and an item table for, e.g., sales orders, is adapted in order to be as close to real-world scenarios as possible.

IoT data is stored in another table with the columns \emph{ID, WORKPLACE\_ID, SENSOR\_ID, DATE,}  as well as columns related to specific sensor measurements, namely \emph{TEMPERATURE\_VALUE, TEMPERATURE\_UNIT, NOISE\_VALUE, NOISE\_UNIT, VIBRATION\_VALUE,} and \emph{VIBRATION\_UNIT}.
As there are only three kinds of sensors, the last columns are specific to these.
When, e.g., a new temperature value is sent and inserted, the columns storing information about the other two sensor types stay empty for that row.

Horizontal integration is achieved using IDs whereas the process of vertical integration makes use of a time-based approach. 
Particularly, a link between sensor data and ERP data can be created as \emph{PRODUCTION\_ORDER\_POSITION} stores the information when a product entered or left a certain workplace. 
Since sensor data also comes with a timestamp, a connection between measurements and workplaces, and thus, between IoT data and products can be established. 

\section{Features and Demo Scenario}
\label{sec:featuresanddemoscenario}

\begin{figure}[!htb]
	\includegraphics[width=\textwidth]{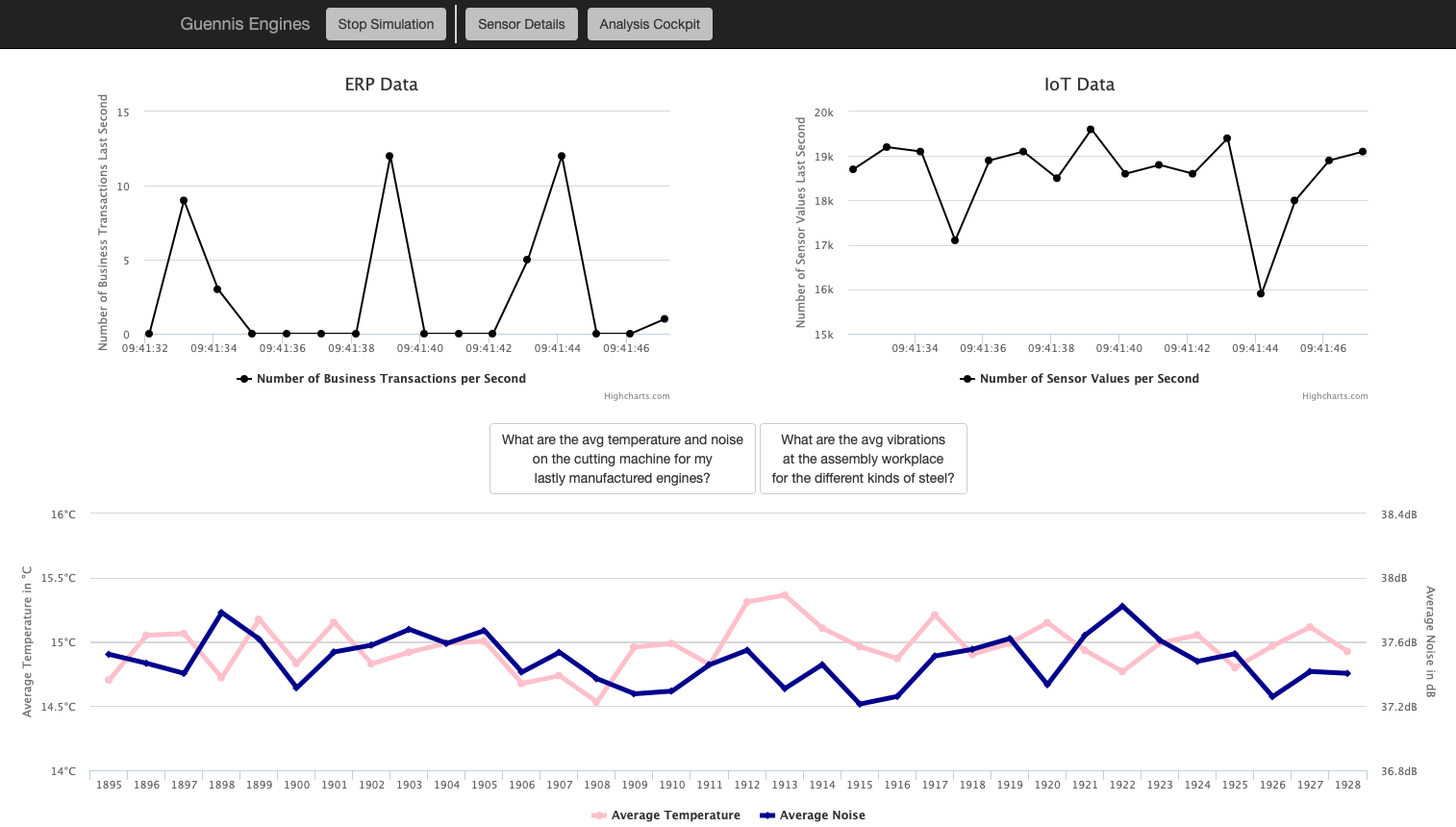}
	\caption{Screenshot a Selected Part of the Demo Application}
	\label{fig:screenshot}
\end{figure}

The demo shown in Figure~\ref{fig:screenshot} allows simulating an industrial manufacturing company, which, from a data perspective, produces business as well as sensor data. 

Ad-hoc queries combining sensor data and business data can be executed. 
All query results are presented in form of a table. 
Furthermore, two sample queries are provided, which answer the questions:

\begin{itemize}[nolistsep]
  \item What are the average temperature and noise on the workplace cutting machine for my recently manufactured products?
  \item What are the average vibrations at the assembly workplace dependent on the supplier?
\end{itemize}

At the top of Figure~\ref{fig:screenshot}, there are three buttons.
The one in the upper-left allows starting and stopping data generation. 
The two diagrams on the top visualize the input rate of business and IoT data. 
The sensor characteristics are defined in a JSON file.
Clicking on the button in the middle opens the sensor config area which allows, e.g., adding certain kinds of sensors to workplaces that produce data at a definable input rate.  
This area is not shown in Figure~\ref{fig:screenshot}.

Below the upper two diagrams, there are two more buttons triggering the execution of one of the two mentioned predefined queries. 
The lower diagram shows the result of the first query, i.e., average temperature and noise values for the lastly manufactured products at the cutting machine.
Not part of Figure~\ref{fig:screenshot} is the result table which presents the raw data  belonging to issues queries as well as the query formulation area, where the predefined queries can be adapted or any ad-hoc query can be inserted and executed.

\section{Conclusion and Future Work}
\label{sec:conclusion}

The presented tool allows experiencing horizontal and vertical integration in scenarios with a real-world character. 
IoT data can be configured and influences on, e.g., performance can be analyzed.
Next to predefined queries that answer valuable questions, any ad-hoc query on the collected data can be executed. 
Results of given queries are visualized in a diagram. 
To the best of our knowledge, the presented demo application is the first of its kind, i.e., a program providing an explorable Industry 4.0 environment with focus on scenarios close to real-world systems and use cases.
Experiments with regard to data integration strategies, data volumes, and resulting impact analysis on query performance can be done easily.
As the code is provided, adaptions and further developments are possible.

%
%
%
\bibliographystyle{splncs04}
\bibliography{mybibliography}

\end{document}